# Spin-Phonon Coupling and Superconductivity in Fe Pnictides

T. Egami[1,2,3], B.-V. Fine[4], D. Parshall[2], A. Subedi[2] and D. J. Singh[3]

[1]*Joint Institute for Neutron Sciences and Department of Materials Science and Engineering, University of Tennessee, Knoxville, TN 37996, USA*

[2]*Department of Physics and Astronomy, University of Tennessee, Knoxville, TN 37996, USA*

[3]*Oak Ridge National Laboratory, Oak Ridge, TN 37831, USA*

[4]*Institute for Theoretical Physics, University of Heidelberg, Heidelberg, 69120, Germany*

**Abstract**

We propose a theory of strong spin-lattice and spin-phonon coupling in iron pnictides and discuss its implications on superconductivity. Strong magneto-volume effect in iron compounds has long been known as the Invar effect. Fe pnictides also exhibit this effect, reflected in particular to the atomic volume of Fe defined by the nearest neighbor atoms. Through the phenomenological Landau theory, developed on the basis of the calculations by the density functional theory (DFT) and the experimental results, we quantify the strength of the spin-lattice interaction as it relates to the Stoner criterion for the onset of magnetism. We suggest that the coupling between electrons and phonons through the spin channel may be sufficiently strong to be an important part of the superconductivity mechanism in Fe pnictides.

## 1. Introduction

Recent discovery of superconductivity in Fe pnictides [1] with the critical temperature $T_C$ up to 55 K [2] caused enormous excitement in the field, for various reasons. First, this is the first non-cuprate family of superconductors with $T_C$ above 40 K. Second, superconductivity appears when the antiferromagnetic (AFM) order is suppressed by doping [3], just as in the cuprates. Third, unlike the cuprates, strong electron correlations are not observed by spectroscopy [4], suggesting the Mott physics may not be a necessary ingredient for the mechanism of high-temperature superconductivity. Finally, there is a large family of similar compounds that show superconductivity, making experimental research less restricted by chemical or materials issues. The field is making surprisingly fast development, partly because of the accumulated experience of working on the cuprates. It is possible that the origin of the superconductivity in this family of compounds may be easier to identify than for the cuprates, and the success of solving this problem hopefully will facilitate understanding of the cuprate problem.

Even though the AFM order is suppressed by doping spins are active in the doped Fe pnictide superconductors. Again, just as in the cuprates strong magnetic excitations, including the so-called resonance peak, are observed by inelastic neutron scattering [5-8]. Core level spectroscopy is consistent with the local Fe moment of about 1 $\mu_B$ [4]. Interestingly the density functional theory (DFT) calculations always predict the magnetic ground state (AFM or incommensurate order) for the experimental lattice constants [9-12]. The phonon dispersions observed by inelastic x-ray scattering are consistent with the DFT calculations only when the magnetic ground state is assumed [13]. Only in the collapsed phase of $CaFe_2As_2$, in which the $c$-axis lattice constant is reduced by as much as 10% compared to the magnetic state, does the material becomes truly spin-degenerate [14]. All these observations strongly suggest that the superconducting samples are locally and dynamically spin-polarized, and show strong dynamic Fe spin fluctuations. Although we do not have precise knowledge of their spin dynamics, judged from the absence of strong quasi-elastic scattering in neutron scattering with energy resolution of 1 meV, the time-scale of fluctuation must not be slower than 1 ps. This result supports the view that spins are involved in the mechanism of superconductivity, for instance though the spin-fluctuation mechanism [15-17].

However, there are many puzzling, important questions that need to be answered before addressing the question of the mechanism: The first puzzle is the effect of doping. In the cuprates doping is necessary for introducing mobile charge carriers, since the parent compounds are Mott-Hubbard insulators. In Fe pnictides, on the other hand, the parent compounds are already metallic, and doping does not appear to change the charge density very much [9]. Rather, the main effect of doping is to suppress the AFM ground state. In the pnictide parent compounds strongly two-dimensional spin fluctuations are observed above $T_N$ [18], just as in the superconducting Fe pnictides [6,8]. However, whereas the LaFeAsO (1111) type compounds are strongly two-dimensional [10], $BaFe_2As_2$ (122) type compounds are much more three-dimensional [19]. The second curious behavior is that the observed magnetic moment on the AFM phase varies from compound to compound, but is always significantly smaller than predicted by the DFT calculations [10-12]. Third, there is an interesting interplay between both the lattice and magnetism [20], and the lattice and superconductivity [21,22]. In this article we focus on the third point, that of the lattice effect. For the purpose of highlighting the essence of the effect we use simple approximations, namely the Landau theory and the Stoner theory, using the results of the LDA calculations as a guide. We argue that through the spin-lattice coupling effect the lattice may play a much larger role than generally acknowledged in determining the properties of Fe pnictides, possibly including even the superconductivity.

## 2. Magneto-volume effect in Fe pnictides

### 2.1 Dependence of Fe moment on the structure

It has long been known that the magnetic moment of transition metals depends on volume [23]. Because of the Pauli exclusion principle the electron kinetic energy of the spin-polarized state is higher for parallel spins if the volume is the same, and volume expansion relaxes the kinetic energy. In some iron alloys thermal volume expansion due to lattice anharmonicity cancels the decrease in volume associated in the decrease in spin-splitting, resulting in zero thermal expansion, widely known as the Invar behavior. The negative or zero thermal expansion is indeed observed for PrFeAsO [24]. The collapsed phase of $CaFe_2As_2$ is a dramatic case of such a magneto-volume effect. This compound shows AFM order below 140K, but with the pressure of 0.4 GPa it undergoes the first order phase transition into a non-magnetic phase with the reduction in volume of 5 % [14].

In Fe pnictides layers of Fe atoms are sandwiched by layers of pnictide such as As or P [1]. Thus if the layer-layer distance of pnictide is changed the magnetic moment of Fe is strongly affected. This coupling of the Fe moment to the pnictide position in the lattice was recognized early by the DFT calculations [9,20]. Fig. 1 shows the calculated dependence of the Fe moment in $Ba(Fe_{0.92}Co_{0.08})_2As_2$ on the separation between the As layer and the Fe layer ($z$). The calculations were done within virtual crystal approximation (VCA) and local density approximation (LDA) with general potential linearized augmented plane-wave (LAPW) method [25], including local orbitals [26]. LAPW sphere radii of 2.2 $a_0$, 2.0 $a_0$, and 2.0 $a_0$, where $a_0$ is Bohr radius, were used for Ba, Fe, and As, respectively. To account for Co doping, an electron number of 26.08 was used for Fe. We used the experimentally reported tetragonal lattice parameters ($a$ = 3.9625 Å, $c$ = 13.0168 Å) [27]. In the calculated result clearly there is a quantum critical point for magnetism (the Stoner condition) near $z_c$ = 1.20 Å as shown in Fig. 1. The local exchange interaction is strong enough to spin-split the band by overcoming the kinetic energy cost only for $z > z_c$. We obtain similar results from the calculation on undoped $BaFe_2As_2$. Compared also with other data [12,20], the relation between $z$ and Fe moment $M$ appears to be rather insensitive to compositions, and the relation shown in Fig. 1 appears to be a nearly universal property of the FeAs triple layer. This must be because the in-plane lattice constant, thus the Fe-Fe distance, is very similar within ± 1 % among many Fe pnictide compounds. Thus the parameter $z$, the As-Fe layer separation, is a good common measure of the magneto-volume effect. For instance, in the collapsed phase of $CaFe_2As_2$ the value of $z$ is 1.23 Å [14], close to the value of $z_c$ in Fig. 1.

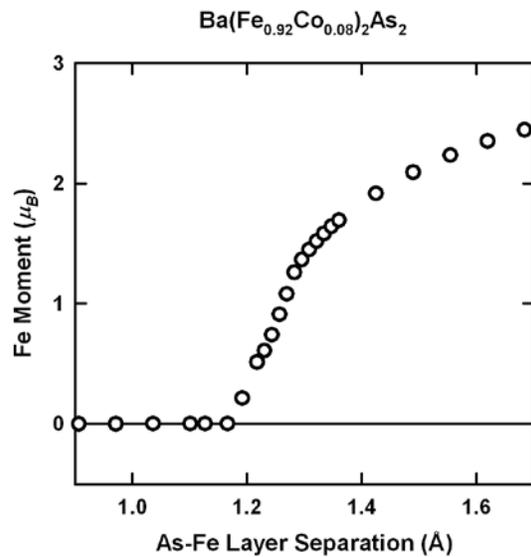

Fig. 1  Fe moment as a function of the As-Fe layer separation in $Ba(Fe_{0.92}Co_{0.08})_2As_2$ calculated by LDA.

## 2.2 Landau theory

Let us develop a Landau-type theory to describe the dependence of the local magnetic moment, $M$, on the As-Fe layer separation, $z$. We may write the magnetic free energy as

$$F_M = AM^2 + BM^4 + F_{s-l} \quad (1)$$

where $F_{s-l}$ is the spin-lattice interaction energy expanded by $z - z_c$,

$$F_{s-p} = \left[\alpha(z-z_c) + \beta(z-z_c)^2\right]M^2 \quad (2)$$

where $\alpha < 0$. $F_{s-p}$ is negative only for $z > z_c$. We retain only the terms with even powers of $M$ because of the symmetry. Following the Stoner condition we set $A = 0$ as discussed below. By minimizing $F_M$ with respect to $M$ we obtain,

$$M^2 = \frac{|\alpha|}{2B}\left[(z-z_c) + \frac{\beta}{\alpha}(z-z_c)^2\right] \quad (3)$$

Fig. 2 gives the fit of this equation to the results shown in Fig. 1. From this fit we obtain $z_c = 1.20$ Å, $\alpha/2B = 19.16$ $\mu_B^2$/Å and $\alpha/\beta = -1.40$ Å. In $CaFe_2As_2$ the QCP is hidden because the nature of the transition is strongly first-order.

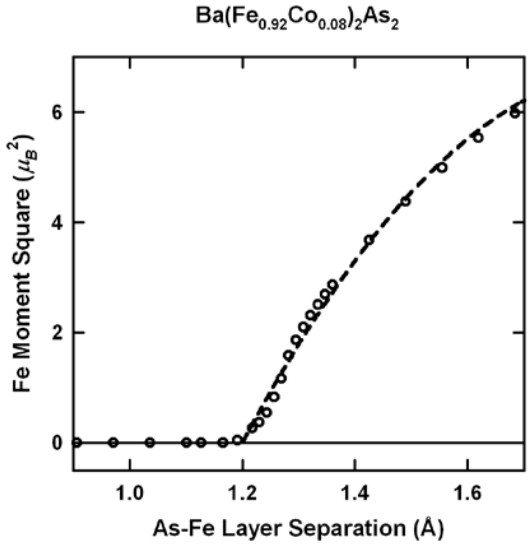

Fig. 2 Square of the Fe moment vs. the As-Fe layer separation. The dashed line is a fit by eq. (3).

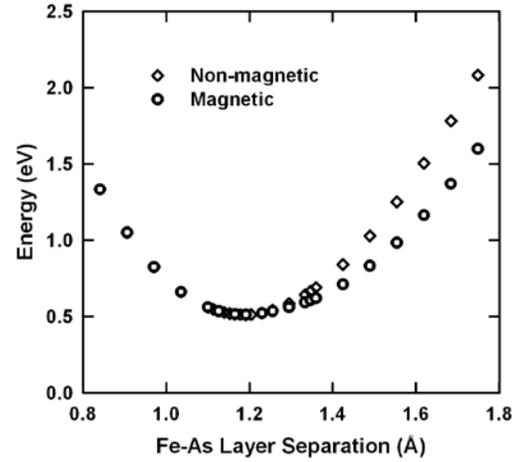

Fig. 3 Electron energy as a function of the Fe-As layer separation calculated by LDA for $Ba(Fe_{0.92}Co_{0.08})_2As_2$, with or without spin polarization, showing phonon softening due to spin polarization.

## 2.3 Phonon softening

We now add the lattice elastic energy in order to consider the phonon softening due to the magneto-volume effect. The phonon to be considered here is the As Raman mode, in which As layers move against each other along the *c*-axis, either toward or away from the Fe layer. The magneto-elastic free energy is

$$F_S(z) = \left[\alpha(z-z_c) + \beta(z-z_c)^2\right]M^2 + K(z-z_1)^2$$
$$= -\frac{\alpha^2(z-z_c)^2}{2B}\left[1+\frac{\beta}{\alpha}(z-z_c)\right]^2 + K(z-z_1)^2 \quad (4)$$

By minimizing $F_S$ with respect to $z$ we obtain the As position, $z_M$,

$$\left.\frac{\partial F_S}{\partial z}\right|_{z=z_M} = -\frac{\alpha^2(z_M-z_c)}{B}\left[1+\frac{2\beta}{\alpha}(z_M-z_c)\right]\left(1+\frac{\beta}{\alpha}(z_M-z_c)\right) + 2K(z_M-z_1) = 0 \quad (5)$$

Then by taking the second derivative we obtain the elastic stiffness renormalized by the spin-lattice interaction,

$$K' = K\left\{1 - \frac{\alpha^2}{2KB}\left[1+\frac{6\beta}{\alpha}(z_M-z_c) + 6\left(\frac{\beta}{\alpha}\right)^2(z_M-z_c)^2\right]\right\} \quad (6)$$

As shown in Fig. 3, allowing spin polarization softens the As Raman phonon frequency by 29% at $z_M = 1.36$ Å. This effect was noted earlier [12], and agrees with the experimental observations [13]. Note that the energy minimum of the DFT calculation systematically underestimates the lattice constant. Thus we obtain, $\alpha = -0.193$ eV/Å $\mu_B^2$, $\beta = 0.137$ eV/Å$^2$ $\mu_B^2$, and $z_1 = 1.32$ Å. The value of $z_1$ is in good agreement with the LDA calculation (1.32 Å after correcting for the systematic underestimation), proving the internal consistency. Thus this theory elucidates how the magneto-volume interaction, eq. (2), can induce magnetization as a function of the As-Fe layer separation, and softening of the As Raman phonon mode.

### 3. Stoner condition

We now turn our attention to the Fe band splitting. For simplicity and clarity of logic we stay in the classical Stoner approximation [28], whereas obviously more complex and accurate DFT calculations can be made. In the Stoner theory the electron energy is given by

$$E = \int_0^{\varepsilon_{F\uparrow}} \varepsilon N_\uparrow(\varepsilon_\uparrow) d\varepsilon_\uparrow + \int_0^{\varepsilon_{F\downarrow}} \varepsilon N_\downarrow(\varepsilon_\downarrow) d\varepsilon_\downarrow + I n_\uparrow n_\downarrow \quad (7)$$

where $N_\uparrow(\varepsilon)$ is the electron density of states for up spins, $n_\uparrow$ is the density of electrons with up spin, and $\varepsilon_\uparrow$ is the energy of an electron with up spin. Magnetization is given by $M = n_\uparrow - n_\downarrow$. If we start with the non-magnetic state and introduce a small spin splitting by shifting the up spin by $d\varepsilon_\uparrow$, the energy change

$$dE = 2N(\varepsilon_F)\left[1 - IN(\varepsilon_F)\right]d\varepsilon_\uparrow^2 \quad (8)$$

which gives the Stoner criterion, $IN(\varepsilon_F) = 1$. Also eq. (9) yields,

$$dE = \frac{1 - IN(\varepsilon_F)}{2N(\varepsilon_F)} M^2 \tag{9}$$

near the Stoner QCP, thus

$$A = \frac{1 - IN(\varepsilon_F)}{2N(\varepsilon_F)} \tag{10}$$

in eq. (1). Therefore A = 0 at the Stoner QCP, justifying the assumption above. We now introduce the effect due to lattice strain, *de*.

$$N(\varepsilon) = N_0(\varepsilon) + N_1(\varepsilon) de + N_2(\varepsilon) de^2 + \ldots$$
$$N_1(\varepsilon) = \frac{dN_0(\varepsilon)}{de}, \quad N_2(\varepsilon) = \frac{1}{2} \frac{d^2 N_0(\varepsilon)}{de^2} \tag{11}$$

$$I = I_0 + I_1 de + I_2 de^2, \quad I_1 = dI/de, \quad I_2 = (1/2) d^2 I/de^2$$
$$\varepsilon_F = \varepsilon_{F0} + \varepsilon_{F1} de + \varepsilon_{F2} de^2 \tag{12}$$

Thus

$$E = E_0 + E_1 de + E_2 de^2 + \ldots \tag{13}$$

By imposing the sum rule that $dn = 0$

$$E_1 = 2\left[\int_0^{\varepsilon_{F0}} \varepsilon N_1(\varepsilon) d\varepsilon - \varepsilon_F \int_0^{\varepsilon_{F0}} N_1(\varepsilon) d\varepsilon\right] + I_1 n_\uparrow n_\downarrow \tag{14}$$

$$E_2 = 2\int_0^{\varepsilon_{F0}} \varepsilon N_2(\varepsilon) d\varepsilon - 2\varepsilon_F \frac{N_1(\varepsilon_F)}{N_0(\varepsilon_F)} \int_0^{\varepsilon_{F0}} \varepsilon N_1(\varepsilon) d\varepsilon$$
$$-2\varepsilon_F \left[\int_0^{\varepsilon_{F0}} N_2(\varepsilon) d\varepsilon - N_1(\varepsilon_F) \frac{\int_0^{\varepsilon_{F0}} N_1(\varepsilon) d\varepsilon}{N_0(\varepsilon_F)}\right] + I_2 n_\uparrow n_\downarrow + \frac{K}{2} \tag{15}$$

By expanding the electronic energy around $z = z_c$, ($de = z - z_c$) and comparing with eq. (5) we obtain,

$$E_1 = 2K(z_M - z_c) \tag{16}$$

$$E_2 = K - \frac{\alpha^2}{2B} \tag{17}$$

## 4. Relation to superconductivity

One of the most intriguing lattice effect on the superconductivity of Fe pnictides is the dependence of the critical temperature, $T_C$, on the geometry of the FeAs$_4$ tetrahedron [21,22]. Data show that $T_C$ is strongly related to the As-Fe-As angle. Because the As-Fe-As angle is directly related to the Fe-As layer separation, $z$, in Fig. 4 we plotted $T_C$ as a function of $z$, using the published results of crystallographic analysis. The results are shown also in Table I, with references for the data. Clearly the behavior above $z_a = 1.4$ Å is different from that below $z_a$.

Above $z_a$ $T_C$ is not much dependent on $z$, whereas there is strong dependence below. It is possible that the systems with $z > z_a$ are regular BCS superconductors, and the enhancement is present only for $z < z_a$. Below $z_a$ there is recognizable correlation between $T_C$ and $z$, except for a few outliers. This correlation supports the idea that $z$ is a universal parameter for the properties of FeAs compounds, including the superconductivity, regardless of the composition.

It is interesting to note that, from eqs. (2) and (3),

$$F_{s-p} = -2BM^4 \qquad (18)$$

Thus we plotted $T_C$ against $M^4$ calculated by eq. (3), not the experimental values of $M$, in Fig. 5. Again strong linear correlation is observed. An obvious implication of this correlation is that indeed magnetism is deeply involved in superconductivity, even though there is no static magnetic order in most of the superconducting samples. However, the involvement of spins in the mechanism may not be limited to the spin-fluctuation mechanism. The linear correlation seen in Fig. 5 implies that the driving energy for spin polarization, eq. (18), resulted in superconductivity instead of magnetism. A possible way that it happens is that the spin-lattice coupling is involved in the superconductivity mechanism through the electron-phonon ($e$-$p$) coupling in the spin-channel, rather than the conventional charge-channel which is weak for the Fe pnictides [45]. The evaluation of the strength of this coupling is in progress.

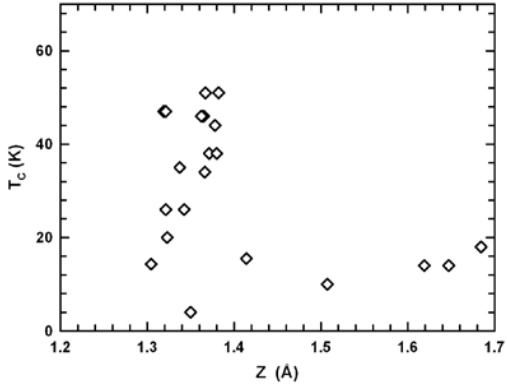
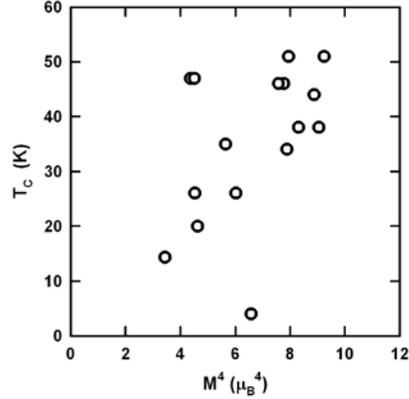

Fig. 4 Superconducting critical temperature, $T_C$, as a function of the As-as a function of the Fe layer separation, $z$.

Fig. 5 Superconducting critical temperature, $T_C$, against $M^4$ calculated by the LDA.

## 6. Conclusions

The conventional electron-phonon ($e$-$p$) coupling through the charge channel is quite small for the Fe pnictide compounds [45]. This led many to conclude that lattice and phonons are irrelevant to the superconductivity of the Fe pnictides. Consequently the spin-fluctuation mechanism is regarded to be the leading mechanism to explain their high $T_C$ [15]. However, the lattice is intimately involved in the magnetism of this compound through the magneto-volume effect, as shown in this paper. This coupling has been known for a long time as Invar effect. The lattice controls the Stoner condition, and thus the onset of spin-splitting. Because the lattice is so intimately involved in magnetism, it is furthermore possible that the *e-p* coupling through the spin-channel is relevant, for instance involving the As Raman phonon mode. The lattice effect may be much more important than generally assumed to the properties of Fe pnictides, including superconductivity.

**Acknowledgements**

The authors are grateful to I. I. Mazin, T. Yildirim, A. Bussman-Holder, N. Mannella and D. J. Scalapino for stimulating and useful discussions. This work was supported by the Department of Energy, Office of Basic Sciences, through the EPSCoR grant, DE-FG02-08ER46528. The work at the Oak Ridge National Laboratory was supported by the Scientific User Facilities Division and by the Division of Materials Science and Engineering, Office of Basic Energy Sciences, Department of Energy.

Table 1   The values of $z$, $T_C$ for various Fe pnictides.  Na-111 means NaFeAs, K-Ba122 means $K_{1-x}Ba_xFe_2As_2$, etc.

| Compound | z (Å) | $T_C$ (K) | Ref. |
|---|---|---|---|
| Na111 | 1.41 | 15.5 | 29 |
| K-Ba122 | 1.37 | 38 | 30 |
| F-La1111 | 1.34 | 26 | 31 |
| F-Nd1111 | 1.37 | 46 | 32 |
| F-Ce1111 | 1.34 | 35 | 33 |
| F-Pr1111 | 1.32 | 47 | 34 |
| F-La1111 | 1.32 | 20 | 35 |
| Pr1111 | 1.32 | 47 | 34 |
| O-Sm1111 | 1.37 | 34 | 36 |
| V-Nd1111 | 1.38 | 51 | 37 |
| Co-La1111 | 1.34 | 14.3 | 38 |
| $Fe(Se_{0.416}Te_{0.584})$ | 1.65 | 14 | 39 |
| $Fe(Se_{0.493}Te_{0.507})$ | 1.62 | 14 | 39 |
| Li111 | 1.51 | 10 | 40 |
| Li111 | 1.68 | 18 | 41 |
| F-Sm1111 | 1.36 | 46 | 42 |
| F-Tb1111 | 1.38 | 44 | 43 |
| V-Nd1111 | 1.37 | 51 | 37 |
| Co-SrFeAsF | 1.35 | 4 | 44 |
| K-Ba122 | 1.38 | 38 | 30 |